\documentclass{aa} 
\usepackage{graphicx}
\usepackage{txfonts}
\usepackage{lscape}
\usepackage{natbib}
\bibpunct{(}{)}{;}{a}{}{,}
\defcitealias{o04}{Paper~I}

\newcommand{\fma}[1]{\mbox{$#1$}}

\newcommand{\unit}[1]{\ifmmode \:\mbox{\rm #1}\else \mbox{#1}\fi}
\newcommand{\mone}{\fma{^{-1}}}

\newcommand{\eg}{{e.g.\/}}

\newcommand{\ie}{{i.e.\/}}
\newcommand{\cf}{{cf.\/}}
\newcommand{\ha}{H$\alpha$}
\newcommand{\hb}{H$\beta$}

\newcommand{\hi}{H~{\sc i}}
\newcommand{\hii}{H~{\sc ii}}

\newcommand{\oii}{[O~{\sc ii}]}
\newcommand{\oiii}{[O~{\sc iii}]}
\newcommand{\siii}{[S~{\sc iii}]}

\newcommand{\caii}{Ca~{\sc ii}}

\newcommand{\kms}{\unit{km~s\mone}}
\newcommand{\cm}{\unit{cm}}

\newcommand{\msun}{\unit{M$_\odot$}}

\newcommand{\flux}{\unit{erg~s\mone\,\cm$^{-2}$}}

\newcommand{\esoig}{ESO\,338-IG04}
\newcommand{\esog}{ESO\,400-G43}

\begin{document}

   \title{Stellar dynamics of blue compact galaxies}

   \subtitle{II. Further indications of a merger in
   ESO\,338-IG04\thanks{Based on observations collected at the European
   Southern Observatory, Paranal, Chile, under observing programme
   65.N-0668.}}

   \author{
          Robert J. Cumming\inst{1} \and 
          Kambiz Fathi\inst{1,2} \and 
          G{\"o}ran {\"O}stlin \inst{1} \and
          Thomas Marquart\inst{3} \and
          Isabel M{\'a}rquez\inst{4} \and
          Josefa Masegosa\inst{4} \and
          Nils Bergvall\inst{3}  \and
	  Philippe Amram\inst{5}
          }

   \offprints{R. Cumming}

   \institute{Department of Astronomy, Stockholm University, SE-106~91 Stockholm, Sweden.\\
              \email{robert@astro.su.se, kambiz@astro.su.se, ostlin@astro.su.se}
          \and
	Instituto de Astrof{\'\i}sica de Canarias, V{\'\i}a L{\'a}ctea s/n, 38200 La Laguna, Tenerife, Spain \and
	Department of Astronomy and Space Physics, Uppsala University, Box 515, SE-751~20
	Uppsala, Sweden \and
	Instituto Astrof{\'\i}sica de Andaluc{\'\i}a (CSIC), C/ Camino Bajo de Hu{\'e}tor 24, 18008 Granada, Spain \and
	Laboratoire d'Astrophysique de Marseille, OAMP, Universit{\'e} de
Provence \& CNRS, 2 Place Le Verrier, 13248 Marseille Cedex 04, France
}

   \date{Received October 25; accepted December 14, 2007}

   \abstract 
{Luminous blue compact
   galaxies, common at $z\sim1$ but now relatively rare, show
   disturbed kinematics in emission lines. } 
{As part of a programme to understand their formation and
   evolution, we have investigated the stellar dynamics of a number of
   nearby objects in this class.}  
{We obtained long-slit spectra with VLT/FORS2 in the spectral
   region covering the near-infrared calcium triplet.  In this paper
   we focus on the well-known luminous blue compact galaxy \esoig\
   (Tololo\,1924--416). A previous investigation, using Fabry-Perot
   interferometry, showed that this galaxy has a chaotic \ha\ velocity
   field, indicating that either the galaxy is not in dynamical
   equilibrium or that \ha\ does not trace the gravitational
   potential due to feedback from star formation.  } 
{Along the apparent major axis, the stellar and ionised gas velocities
for the most part follow each other. The chaotic velocity field must
therefore be a sign that the young stellar population in \esoig\ is
not in dynamical equilibrium. The most likely explanation, which is
also supported by its morphology, is that the galaxy has experienced a
merger and that this has triggered the current starburst.
   Summarising the results of our programme so far, we note
   that emission-line velocity fields are not always reliable tracers
   of stellar motions, and go on to assess the implications for kinematic
   studies of similar galaxies at intermediate redshift.  }
   {}
\maketitle
%
%

\section{Introduction}

In the early universe, luminous, blue, compact galaxies (LBCGs) were
much more common than they are now. Compact, emission-line-dominated
galaxies with blue colours, small radii and disturbed morphologies
\citep[\eg][]{koo,g96} may have accounted for nearly half of all
star formation \citep{g97} at redshifts between 0.4 and 1.  Since then,
the contribution from such galaxies has dropped
by a factor of about ten \citep[ and references therein]{Werk,g97,Garland}.  

LBCGs at all redshifts show both disturbed morphology and often also
disturbed gas kinematics \citep{o99,o01,gdp,Pisano,Bershady,Puech}.
These have been interpreted as signs of a merger history \citep{o01},
and disturbed velocity fields are reproduced in models of merging
galaxies \citep{Jesseit,Kronberger}.  There is, however, no {\it a
priori} reason to assume that the stars and the emitting gas in these
galaxies share the same kinematics. Moreover, \citet{o01}, studying
the \ha\ velocity fields of \citet{o99}, found LBCGs in their sample
with apparent dynamical mass deficiencies that could be attributed to
non-equilibrium gas kinematics.  Their results imply that this may be
the rule rather than the exception for the subclass of very luminous
blue compact galaxies.  Such discrepancies could arise either because
the gas is not in equilibrium with the gravitational potential of the
galaxy, or because the gas motions are not solely in response to the
gravitational potential. This second possibility, which could be due
to feedback in the form of outflows, bubbles and superwinds, called
for an investigation of the {\em stellar} kinematics of BCGs. Do the
gas and stars in these galaxies move in concert or independently of
one another?

This issue is important for assessing kinematic studies of
star-forming galaxies at high and intermediate redshift, where
velocity fields computed from emission lines may be the only
observational way to probe of the gravitational
potential. \citet{kg00} have compared slit spectra of a sample of 22
local late-type galaxies, among them the LBCG He\,2-10, in emission
lines (\oii, Balmer lines and \hi\ 21 cm) and
\caii\ H and K.  They concluded that stellar velocities and velocity
dispersions generally agreed well with those of the ionised gas, and
the agreement was particularly good for He\,2-10. Moreover, \citet{tm81}
have described a relation between velocity dispersion and \hb\
luminosity in extragalactic giant \hii\ regions, which has been used
to estimate masses \citep[\eg][]{g96,o01}. \citet{melnick87} concluded
that the relation is most likely determined by the virial motions of
young stars.

To investigate the relationship between gas and stellar
kinematics, we have collected kinematic data for a number of LBCGs.
Our tactic has in each case been to trace the velocity field of the
stars using the infrared triplet of Ca~{\sc ii} and compare with the
velocity field seen in emission lines such as [S~{\sc iii}]
$\lambda$9069 and the Paschen series of H~{\sc i}.  The first results
of this program were presented in \citet[ hereafter Paper~I]{o04},
where we used long-slit spectra to investigate the stellar and gas
kinematics of \esog.  In \citet{m07} we used integral field
spectroscopy to map the kinematics of He\,2-10 in both emission lines
and stellar absorption lines.  In future we will present similar
analyses for Haro~11 (ESO\,350-IG38), ESO\,480-IG12 and II\,Zw\,40.
In this paper we present new long-slit spectra of \esoig,
which has a disturbed velocity field in \ha\ \citep{o01}. In the light
of our results, we also take the opportunity to summarise what we
currently know about the stellar and gas motions in LBCGs.

\section{Observations and data reduction}

We have carried out long-slit spectroscopy with FORS2 on the 8-m
telescope UT2/Kueyen at the VLT.  \esoig\ was observed between 2000
July 03 and 2000 August 03 using the Tektronix 2048$\times$2048-pixel
CCD.  
\begin{table}

\caption{Log of observations}\label{tab-observations}
\vspace{\baselineskip}

\begin{tabular}{@{}lllll}

Date &  Object  & Seeing   & Slit & Total integration \\
(UT) &         & (arcsec) & PA ($^\circ$) & time (s) \\
\hline
2000 07 03 & \esoig & 0.7-1.2 & 90$^{\circ}$   & 8100 \\
2000 07 04 & \esoig & 0.9-1.2 & 160$^{\circ}$   & 1800 \\
2000 07 29 & \esoig & 0.6-0.8 & 90$^{\circ}$   & 4200 \\
2000 08 02 & \esoig & 1.0-1.4 & 90$^{\circ}$   & 1200 \\
2000 08 03 & \esoig & 0.4-0.7 & 90$^{\circ}$   & 3600 \\
2000 08 03 & \esoig & 0.4 & 160$^{\circ}$   & 1800 \\
\hline
\end{tabular}

{\footnotesize \raggedright Notes: Wavelength coverage 
was 7850-9230 \AA.   
The spectral resolution, measured from the width of
  unresolved extracted sky emission lines, was 67$\pm$3 km~s$^{-1}$ 

}

\end{table}

\subsection{\esoig}

   \begin{figure}
   \centering
   \includegraphics[width=7.5cm]{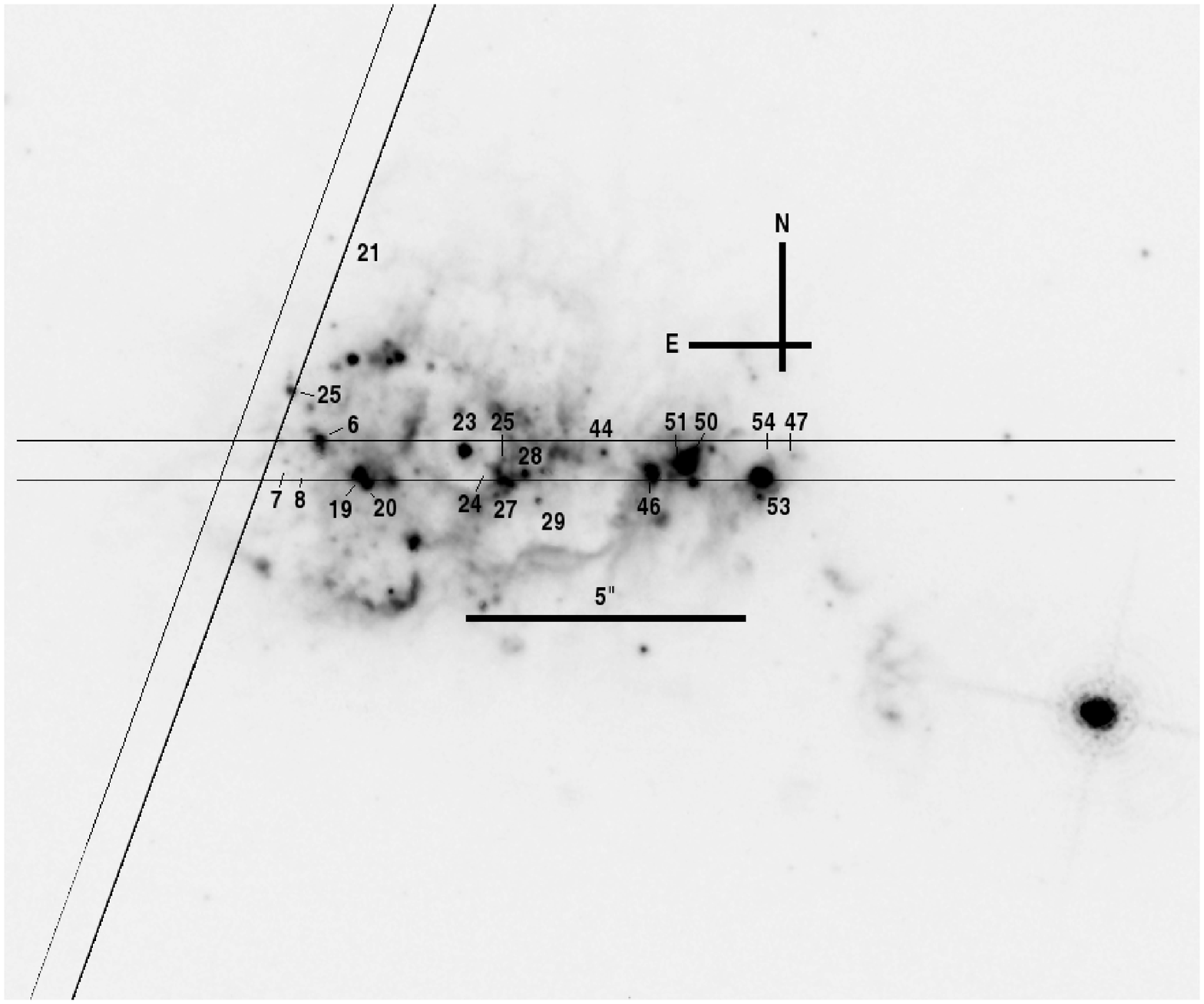}

   \caption{Positions of slits PA 90$^\circ$-270$^\circ$ and
   160$^\circ$-340$^\circ$ superimposed on an HST narrow-band \ha\
   (F656N) image of \esoig\ taken from \citet{Hayes}.  North is up and
   east to the left.  The positions of prominent super star clusters
   are marked, numbered according to \citet{OBR}. All the clusters
   belong to their inner sample, except 21, 25 and 47, which belong to
   the outer sample. The bright object to the lower right is a
   foreground star. The field of view here is 22\arcsec\ by
   18\arcsec, corresponding to 3.9 kpc by 3.25 kpc at the distance of
   \esoig\ \citep[37.5 Mpc;][]{OBR}.}

              \label{f-eso338-slits-Ha}%
    \end{figure}

%
%
%
%
%
%
%
%
%
%
%
%
%
%
%
%
%
%

For our long-slit spectroscopy, we used a slit of width 0.7\arcsec\ at
two different positions for \esoig.  Figure 
\ref{f-eso338-slits-Ha} shows the slit positions relative to the
the star clusters and the nebular
emission in the galaxy.  One slit was aligned along the major axis of
the galaxy, at position angle 90$^\circ$-270$^\circ$.  At this
position we integrated for a total of 18000~s, divided into 1200- and
900-second exposures.  The other slit was chosen to investigate the
steepest velocity gradient seen in \ha\ data for the galaxy \citep[
see also Figure \ref{f-eso338-cigale}]{o99}, at position angle
160$^\circ$-340$^\circ$ and centred 3.7\arcsec\ east of the central
concentration (cluster \#23 in the inner sample of \citealt*{OBR}).  At
this position, however, we obtained only 4$\times$900-s integrations
due to time constraints.  We used the G1028z grism, which gave a
spectral resolution of about 4500 (67$\pm$3 \kms, as measured from
unresolved sky lines) near the \caii\ triplet.

We followed a standard reduction procedure using {\sc
iraf}\footnote{IRAF (Image Reduction and Analysis Facility) is
distributed by the National Optical Astronomy Observatories, which are
operated by the Association of Universities for Research in Astronomy,
Inc., under cooperative agreement with the US National Science
Foundation.}. We bias-subtracted and flat-fielded the science frames
in the usual way, using dome flats to create a mean flat field frame.
Then we removed by hand cosmic ray hits that lay close to the lines we
are most interested in, since we found that not all cosmic rays
were properly removed by median-filtering the final registered
spectra. Wavelength calibration was carried out using the OH sky
lines, with wavelengths from \citet{o92} and \citet{o96}.

The resulting long-slit spectra were combined, shifting first so that
the galaxy centroid along the slit was the same in each case.  While
the spectra were taken under a variety of conditions, and some observations were affected by poor
atmospheric transparency, the seeing was generally good (Table
\ref{tab-observations}).  We tested
different averaging methods and excluding the poorer data, and found
that the best signal-to-noise ratio in the final result was obtained
by median-filtering all the spectra and weighting according to detected
counts in the continuum at 8610-8700 \AA.  For this we used the task
{\sc scombine} in {\sc iraf}.  The resulting wavelength-calibrated,
two-dimensional spectrum was used for the subsequent analysis.
%
%
%
%
%
%
%
%
%
%
%
%
%
%
%
%
%
%
%
%
%
%
%
%
%
%
%
%
%
%
%
%

\subsection{Extraction and flux-calibration}

From our final spectra for each slit, one-dimensional
extractions were made at intervals in the spatial direction (3 detector rows, or 0.75\arcsec). 
The sky background was subtracted from the \esoig\ spectra at this
stage, linearly interpolating between regions free of galaxy
emission on each side of the galaxy. For this we used patches $\sim$1
arcmin wide displaced by 50\arcsec\ from the centre..  The extracted
spectra were in turn flux-calibrated by comparison with spectra taken
of the standard stars LTT 7379 and LTT 7987 using the same set-up.

While we have not corrected our spectra for telluric absorption, the
region containing the \caii\ triplet is relatively free from such
features. In our analysis we have also avoided the region longward of
9000 {\AA} where the H$_2$O bands are strong \citep[see for example Figure
16 in][]{m00}.  The emission lines Pa~11, Pa~10 and
\siii\ $\lambda$9069 do lie in this region, but our iterative
procedure for Paschen-subtraction is unlikely to be affected by
telluric features.

\subsection{Subtracting emission lines}
\label{pasub}

\begin{table}

\caption{Template stars for cross-correlation analysis.}
\label{tab-templates}
\vspace{\baselineskip}

\begin{tabular}{@{}ll}
\hline
Star & Type/class \\

\hline

HD\,164349  &K0 {\sc ii-iii} \\
HD\,172365  &F8 {\sc i}b-{\sc ii}  \\
HD\,174947  &K1 {\sc i}B \\
HD\,174974  &K1 {\sc ii} \\
HD\,175751  &K2 {\sc iii} \\
HD\,179323  &K1 {\sc i}  \\
\hline
\end{tabular}

{\footnotesize \raggedright Note: Spectral types and luminosity classes are taken from the VizieR database \citep{VizieR}. }
\end{table}


%
%

%
%
%
%
%

%
%
%
%
%
%
%
%
%
%
%
%
%
%
%
%
%
%
%
%

%
%

%
   \begin{figure}
   \includegraphics[angle=270,width=.49\textwidth]{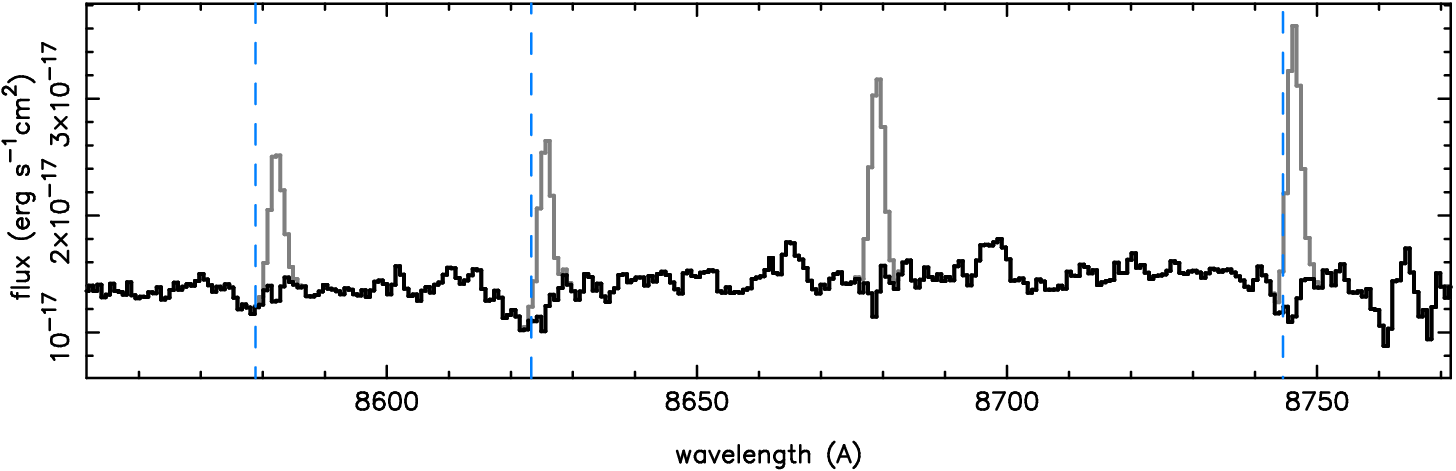}\\
   \includegraphics[angle=270,width=.49\textwidth]{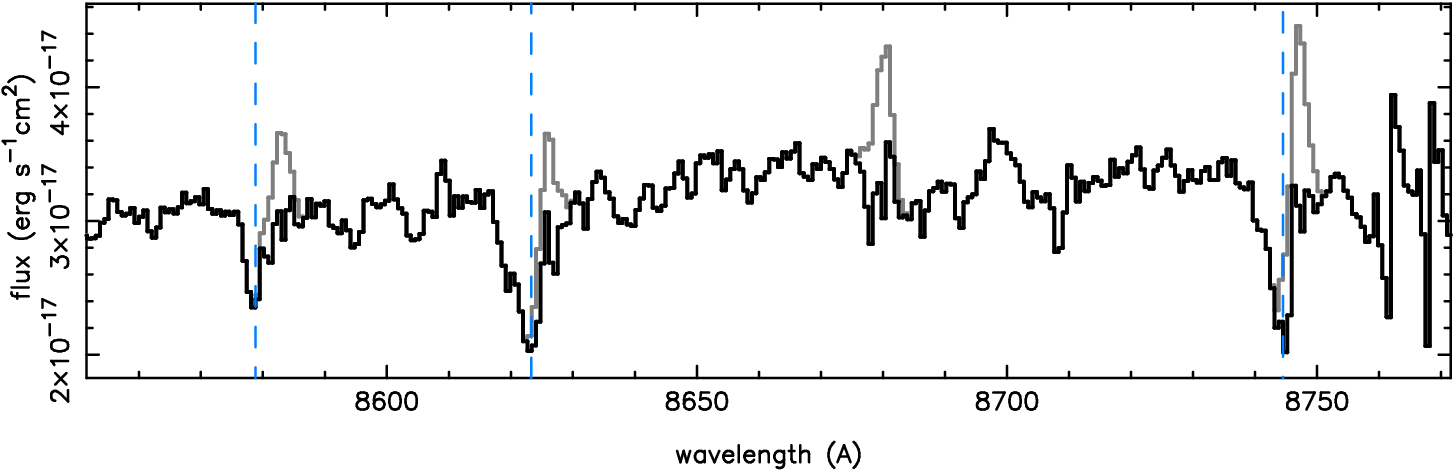}\\
   \includegraphics[angle=270,width=.49\textwidth]{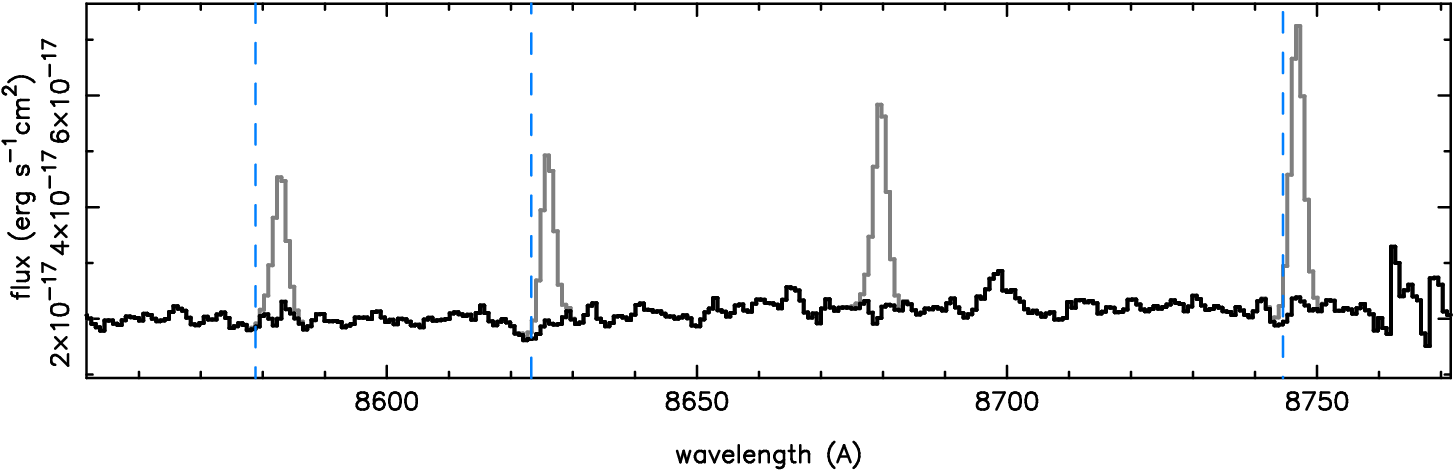}
      \caption{Three examples of Paschen-line subtraction from the
      spectrum of \esoig\ around the calcium triplet, for
      extractions with different signal to noise ratio and different
      quality of subtraction, using the less affected blue wing of each line as a guide. These subtractions, for the slit at PA 90--270\degr, were classed as high
      (top; position $-$2\arcsec.7), medium (middle; +1\arcsec) and low (bottom; $-$1\arcsec.2).  Positions are relative to the point where the two slits cross; see section \ref{results}. The subtracted
      spectrum is plotted over the original spectrum (grey).  Vertical
      dashed lines mark the approximate expected position of the centres of the
      three calcium triplet lines. }  
	\label{f-paschen-sub}
      \end{figure}

Our aim was to compare the extracted galaxy spectra around the \caii\
triplet with the spectra of template stars of different types (Table
\ref{tab-templates}).  The spectra are however dominated by narrow
emission lines, primarily the H~{\sc i} Paschen series. To remove
these lines and achieve a clean spectrum of the galaxy's stars alone,
we model and subtract the Paschen series lines. We have described our 
Paschen-subtraction technique in  Paper~I. For \esoig, where the
emission lines are stronger relative to the stellar absorption lines,
we refined the Paschen-subtraction procedure.

We measured the fluxes of the strongest Paschen lines from our
spectra, excluding those which overlap with the \caii\ triplet,
and compared with the predicted strengths for Case B recombination at
a temperature of 10$^4$ K and density 100 cm$^{-3}$ given by
\citet{o89}.  Within the likely errors of our measurements, the lines
follow the predicted values. This is consistent with the low
extinction values used for these galaxies by \citet[$E_{B-V} < 0.1$;\
][]{OBR}, and we have therefore made no extinction correction to
our data or models.  The Paschen decrement does not vary
significantly across our slits, indicating no significant variation in
extinction.

An initial, first-guess model profile for each line was constructed by
fitting a number of Gaussians to the high signal-to-noise \siii\
$\lambda9069$ line. This model profile was then shifted to match the
measured redshifts of the strongest Paschen lines in the spectrum,
creating a model spectrum of the Paschen series from Pa~9 to Pa~20. We
scaled this spectrum to match the strongest clean Paschen lines
(Pa~10, Pa~11, Pa~12 and Pa~14) and finally subtracted it from the
observed spectrum (Figure \ref{f-paschen-sub}).  Thereafter we
adjusted the width, redshift and strength of the lines, iterating in
order to produce plausible subtractions.  In practice this meant
minimising the subtraction residuals around the clean Paschen lines
and ensuring that the red wing of each corrected \caii\
absorption feature was plausibly consistent with its less
contaminated blue wing.

We judged the reliability of each subtraction qualitatively, dividing
the subtractions into three classes of high, medium and low quality. In
particular, where the risk of over- or undersubtraction was large and
the emission lines were particularly strong, the low-quality
classification is chosen.  In our subsequent analysis, we have only
included those with high or medium quality. Figure \ref{f-paschen-sub}
shows examples of each level.

Finally, we removed sky subtraction residuals and the remaining strong
emission line O~{\sc i} $\lambda$8446 
by linearly interpolating across it. Figure \ref{f-paschen-sub}
shows examples of prepared spectra.

   \begin{figure} \centering
   \includegraphics[width=.49\textwidth]{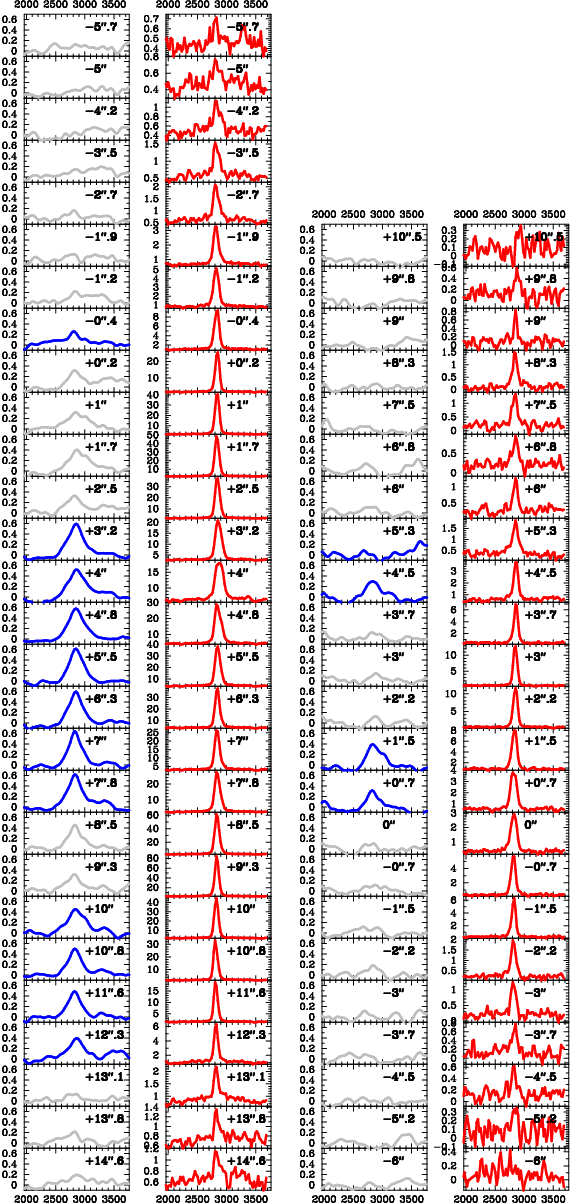}

   \caption{Cross-correlation functions with respect to
   HD\,164349 (blue/darker and grey/lighter profiles; first and third
   panels) and [S~{\sc iii}] profiles (in red; second and fourth
   panels) for \esoig.  The left two boxes are for slit PA
   90$^\circ$-270$^\circ$ and those on the right are for PA
   160$^\circ$-340$^\circ$.  Positions in arcsec relative to the
   crossing point of the two slits are marked next to each profile. 
   CCFs are plotted in grey where the Paschen subtraction quality is
   low, or where the peak of the CCF is less than 0.25.  Units in each
   plot are (abscissae) velocity in \kms\ and (ordinates) CCF value or
   flux in units of $10^{-17}$ \flux.  } \label{f-eso338-ccf} 

   \end{figure}

\subsection{Deriving the stellar kinematics}

To ensure robust measurements of the stellar velocity field, we have
used two independent methods, cross-correlation with template stars (section
\ref{sec-ccf}) and penalized pixel-fitting (pPXF; section \ref{sec-ppxf}). 

\subsubsection{Method 1: Cross-correlation with template stars}
\label{sec-ccf}

We used the {\sc iraf} task {\sc fxcor} to cross-correlate the
prepared spectra with our six template stars on the region containing
the \caii\ triplet.  For this we used only the region 8377-8948 \AA,
\ie\ only Paschen-subtracted part of the spectrum, as far red as the
start of the stronger telluric absorption bands at longer wavelengths.
To ensure that strong signals from pixel-to-pixel noise and
low-frequency undulations in the spectra did not affect measurements
of the cross-correlation function (CCF) peak, we experimented with a
ramp filter defined in {\sc fxcor}. In this we follow \citet{TD}
and \citet{Wegner99}. We found that the best correction for
pixel-to-pixel noise resulted from a ramp rising from zero to unity at
wavenumber 15 (wavenumbers here correspond to power at different
numbers of spectral pixels in the prepared spectrum).  For a good
match between the low-frequency behaviour of the template and galaxy
spectra, we found we needed the ramp filter to fall to zero between
wavenumbers 1000 and 1100.

The resulting CCFs are shown in Figure
\ref{f-eso338-ccf}.  
We measured the width and velocity of each peak by fitting a Gaussian
on the region above 60 percent of the peak height.

For measuring the heliocentric velocity, we required accurate radial
velocities for the template stars (Table \ref{tab-templates}).  We
used catalogue radial velocities from the VizieR database
\citep*{VizieR} and for consistency cross-correlated between pairs of
template stars to determine better radial velocities for some of them.
To do this we adjusted the least precisely determined velocities in
turn, in order to minimise the sum of the squares of the differences
between the expected and measured velocity difference in each pair.

For errors in heliocentric velocities, we have used the errors
calculated by {\sc fxcor}.

To be able to measure the velocity dispersion $\sigma$, we empirically
determined the relationship between the width of the peak of the
cross-correlation function and the value of $\sigma$.  To do this we
broadened the template star spectra with Gaussians of various widths
(\ie\ known $\sigma$), and compared the resulting  velocity width
$v_{\rm FWHM}$ of the cross-correlation peak with the value of
$\sigma$ \citep{nw95, HFa, HFb}.  To avoid systematic errors in this
process, we used the same procedure here described for similar
measurements in \citet{o07}, adding noise to match the value of the
CCF peak.  This ensures that the relation between $v_{\rm FWHM}$ and
$\sigma$ is not affected by mismatches in data quality.

The velocity curve determined from cross-correlations is independent
of the choice of template star.  The mean values between +3.2\arcsec\
and +7.8\arcsec, for example, all agree to within the errors with a
velocity of 2833.5 \kms.  The derived velocity dispersions, on the
other hand, are much larger for the F8 {\sc i}b-{\sc ii} template star
HD\,172365 (mean value $\sigma=114\pm9$ \kms, compared to a mean of
$64\pm4$ \kms\ for the others). This star has broader CaT lines than our
other template stars, and the implication here is that our technique
for calibrating velocity dispersions breaks down when line widths in
the object and template spectra are comparable. We do not include
HD\,172365 in our subsequent analysis.

For our velocity dispersion measurements, we found that the dominant
source of uncertainty, particularly where the signal-to-noise ratio was
low, was the identification of the CCF peak and the Gaussian fit to it.
To quantify the errors introduced in this way to each measurement of
$v_{\rm FWHM}$, we followed the same procedure as \citet{o07}, 
investigating the range of plausible Gaussian fits to the CCF peak. We
then used the standard error in these measurements as an estimate of
the error. In contrast to \citet{o07}, we found that the errors found
in this way are larger than those introduced by the scatter in the
$v_{\rm FWHM}-\sigma$ calibration relation.  The errors were then
propagated through the calibration procedure described above.  This
method is not rigorous but gives, we believe, acceptable estimates of
the error in~$\sigma$.  

For some positions with poor signal-to-noise in the stellar spectrum,
the cross-correlation function peaks at values too low to provide
reliable estimates of either the velocity or the velocity
dispersion. We have therefore defined a cut-off for acceptable
cross-correlations at CCF value 0.25.

\subsubsection{Method 2: Penalised pixel-fitting}\label{sec-ppxf}

We have also derived the stellar kinematics using the penalised pixel
fitting (pPXF) method developed by \citet{CE04}.
In this case we fitted the lines in real pixel space, since it is easier
to mask the spectra to remove bad pixels or regions contaminated by
emission lines. We made use of a large library of high resolution
observed and model spectra \citep{Cenarro01a,Cenarro01b}. We
created linear combinations of these template spectra to match with the
galaxy spectrum and fit the stellar kinematics. The parameters ($V$,
$\sigma$, $h_3$, $h_4$) were fitted simultaneously, with an adjustable
penalty term added to the $\chi^2$ to bias the solution towards a
Gaussian shape when the higher order terms (skewness $h_3$ and
kurtosis $h_4$) are unconstrained.

To make the pPXF routine deliver the best template spectrum quickly, we picked out a restricted set of 17 stellar spectra 
from the Cenarro library. These spectra were carefully chosen to cover
a wide range of stellar types: dwarfs, giants, and
supergiants, spectral classes O9~{\sc v} ($T_{\rm eff}$=36\,300 K) to
M6~{\sc v} ($T_{\rm eff}$=3720 K), with metallicities ranging from
[Fe/H]=$-$2.25 to 0.13 solar.  The best matching combined
spectrum can thus represent regions with distinctly different stellar
populations.  However, since there is more than one combination of
stellar mixes that can reproduce the absorption features in a galaxy
spectrum, the weights assigned to individual stellar spectra could not be
used to carry out a population analysis for our galaxies.

We investigated the robustness of the kinematic parameters delivered
by pPXF by changing the input stellar library, masking out the red
wings of the CaT lines which are where imperfections in the
Paschen-line subtraction could change the derived stellar kinematic
parameters, and finally by applying a higher weight to the stronger
absorption features. We found that in all these cases, we were still
able to derive similar velocities and velocity dispersions.

%
%
%
%
%
%
%
%
%
%
%
To estimate errors for the pPXF kinematic parameters, we applied the
bootstrap Monte Carlo method \citep[Chapter 6 of][]{press}. We ran 300
simulations, varying the input stellar library, both by adding and
subtracting template stars and by using one star at a time.  We
found that the derived kinematical parameters for the full library of 17
template stars 
with parameters
derived for a subset of these or individual stars,
all fall within $\pm 30 $ \kms\ of the adopted values.  This therefore
gives an upper limit for the errors. Additionally, we found that
for the noisier spectra, the pPXF method can produce a spurious fit to
the spectra with weak narrow \caii\ lines which clearly cannot be
justified by the data. We have flagged these points and do not include
them in our analysis.

\subsection{Systematic errors}

Our results from cross-correlation and pPXF agree well, and deliver
comparable stellar velocities and velocity dispersions. This indicates
that the kinematics in \esoig\ can be derived in a robust
way from our prepared spectra.  Both cross-correlation and pPXF
deliver comparable stellar velocities, and the choice of stellar
template is not important as long as the star is a late-type giant or
supergiant.

A remaining source of systematic error is our Paschen-subtraction
procedure. To quantify this, we scaled the flux of each model spectrum
to produce plausible over- and under-subtractions. We then redetermined
the kinematics using pPXF and examined both the quality of the fit
($\chi^2$) and the resulting changes in the derived velocity and
velocity dispersion. For all our spectra, we found the minimum of
$\chi^2$ to fall within 10\% of the adopted strength of the Paschen
subtraction. This corresponds to systematic errors in both velocity
and velocity dispersion of at most 10 \kms. Moreover, we see no
systematic trend towards over- or under-subtraction, giving us
confidence that our method provides a reliable correction for \hi\
emission at each position.

\subsection{Deriving the gas kinematics}
\label{sec-gas}

Emission-line velocities, widths and fluxes were measured from
Gaussian fits to the lines carried out interactively using the IRAF
task {\sc splot}. The velocity dispersion in each emission line was
calculated by subtracting the instrumental width in quadrature from
the measured line width.  Errors were estimated from the scatter in
the measured values for lines of different measured flux. For the
Paschen lines we calculated weighted means for the velocities and
widths using all the available lines at each position.

\section{Results}
\label{results}

In the following we quote positions in arcsec along our slits
relative to the point where the two slits cross (Figure
\ref{f-eso338-slits-Ha}). Positive values for the two slits are
towards the west (PA 270\degr) and north (PA 340\degr),
respectively.

\begin{figure*}
   \centering
   \includegraphics[angle=-90,width=.9\textwidth]{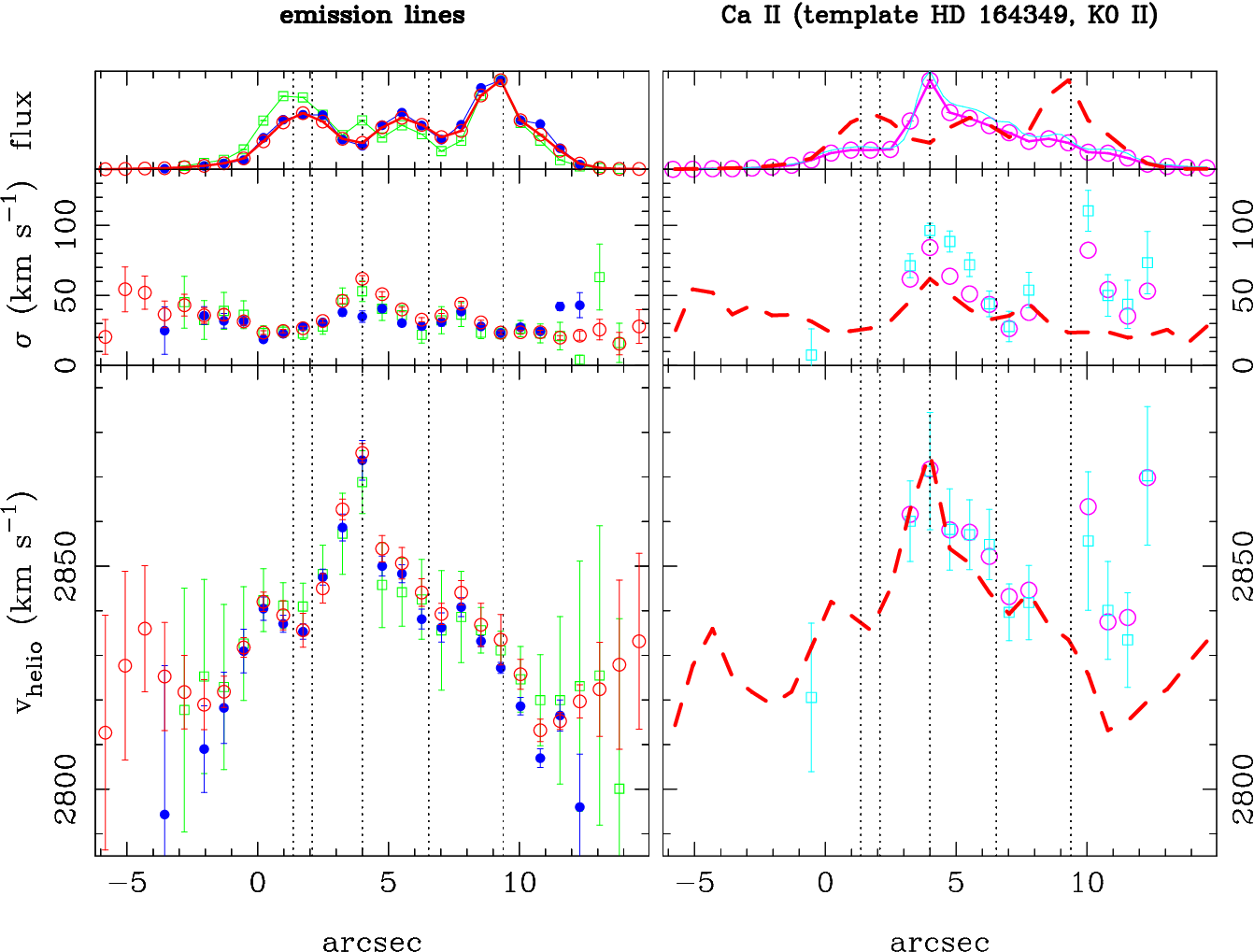}

   \caption{Variation along slit PA 90$^\circ$-270$^\circ$ of
     (from top to bottom) continuum/line flux, velocity dispersion $\sigma$ and heliocentric
     velocity, for the gas (left) and stars (right) in
     \esoig.  Zero on the abscissa is defined as the point where our two
     slits cross, 4\arcsec\ east of cluster \#23.  The left panel
     shows the emission lines \siii\ $\lambda$9069 (red open circles),
     \hi\ Paschen lines (weighted means of all measured lines; blue filled circles), and O~{\sc i} $\lambda$8446
     (green squares).  The flux scales show (left) the normalised strengths of the emission lines  and  (right) the \siii\ line flux compared to the mean continuum value around
     the calcium triplet. 
     The dotted vertical lines mark the positions along the slit of
     (from left to right) clusters 6, 19, 23, 44 and 53 from the inner
     sample of \citet{OBR} (see also Figure \ref{f-eso338-slits-Ha}).
     The right panel shows results of cross-correlation with the K0
     bright giant HD\,164349 (blue squares) and pPXF results (magenta open
     circles).  Cross-correlation measurements of velocity dispersion
     and heliocentric velocity are only shown where the CCF peak is
     greater than 0.25. CCF and pPXF results are not shown where the
     Paschen-subtraction quality was deemed to be low (see Figure
     \ref{f-paschen-sub}).  The thick dashed line in the right panel
     repeats the \siii\ results from the left panel for comparison.
     At the distance of \esoig, 10\arcsec\ corresponds to 1.8 kpc.}

      \label{f-rotcurve-eso338-1}
   \end{figure*}

\begin{figure*}
   \centering
   \includegraphics[angle=-90,width=.9\textwidth]{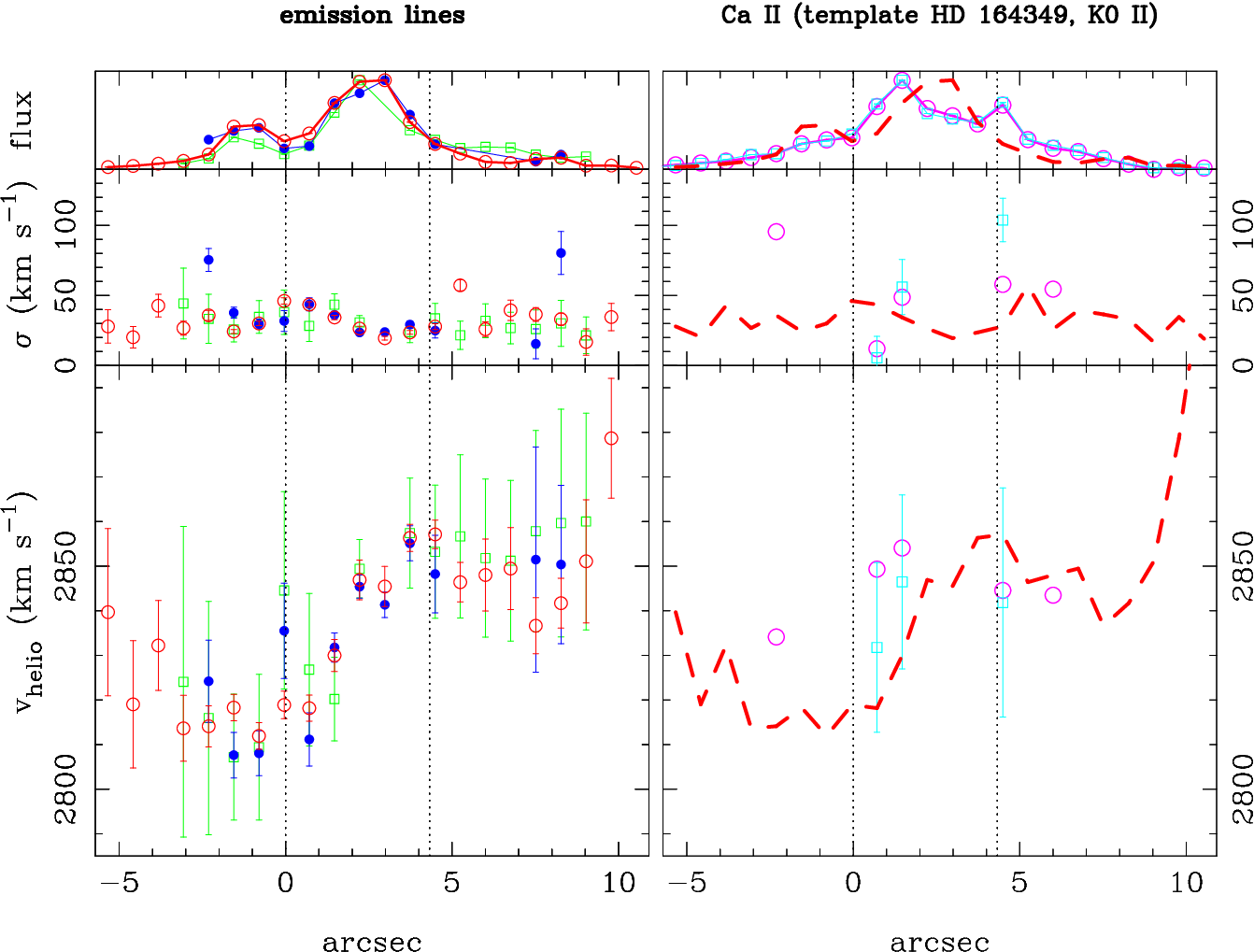}

      \caption{Same as Figure \ref{f-rotcurve-eso338-1}, but for
      \esoig\ at PA 160$^\circ$-340$^\circ$.  The dotted
      vertical lines in this case mark the positions along the slit of
      clusters 25 (left) and 21 from the outer sample of \citet{OBR}
      (see Figure \ref{f-eso338-slits-Ha}).}

         \label{f-rotcurve-eso338-2}
   \end{figure*}

Figures \ref{f-rotcurve-eso338-1} and \ref{f-rotcurve-eso338-2} show
the measured velocities and velocity dispersions plotted against
position along the two slits in \esoig.   The plots show
results from both the cross-correlation and pPXF methods and also the 
spatial variation of the continuum level near the CaT lines (for
the stars) and emission line flux (for the gas).

\begin{figure}
   \centering
   \includegraphics[width=.49\textwidth]{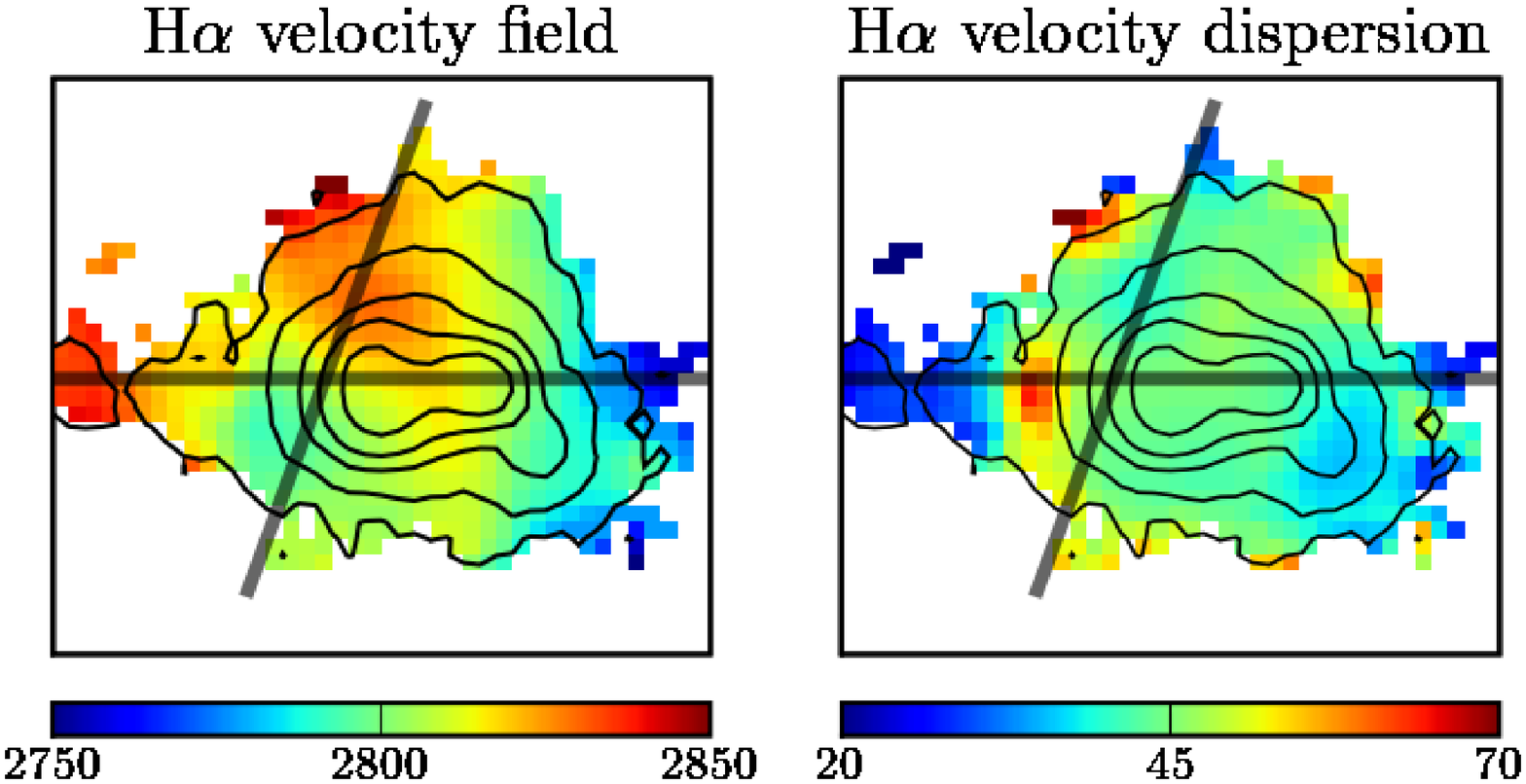}

   \caption{Velocity field (left) and velocity dispersion (right)
   of \esoig\ from Fabry-Perot measurements of \ha, as measured by
   \citet{o99}. North is up and east to the left. The colour scales in
   \kms\ are indicated below each panel. The black contours correspond
   to the \ha\ flux in arbitrary logarithmic units. Overlaid in grey
   are the positions of our slits (\cf\ Figure
   \ref{f-eso338-slits-Ha}). Each frame is 36.4\arcsec\ wide, and one pixel corresponds to 0.91\arcsec\
   on the sky. }
   \label{f-eso338-cigale}%

\end{figure}

Along the slit at PA 90$^\circ$-270$^\circ$, the emission lines show
close to identical velocity curves, and show the same
features as the \ha\ observations of \citep[ Figure
\ref{f-eso338-cigale}]{o99}. The velocity field has a span of about 60
\kms\ with local minima at $-$2\arcsec\ and +11\arcsec, and a barely
resolved (1.5\arcsec-wide) local maximum with amplitude $\sim$30 \kms\
at the position of cluster \#23 (+4\arcsec).

Although the stellar velocity field is much less well-determined than
that of the gas, some trends are nevertheless evident. Just as for the
emission lines, no strong velocity gradient is detected. Indeed, the
data are marginally consistent with a flat rotation curve with
velocity 2860 \kms. A more interesting null hypothesis, however, is
that the stars show the same velocity field as the gas. Between
+3\arcsec and +8\arcsec the agreement is clear from Figure
\ref{f-eso338-cigale}.  Between +10\arcsec\ and +13\arcsec, however, the stars
and gas differ at the 3-$\sigma$ level: the mean stellar velocity is
2846$\pm$8 \kms\ (taking the mean of all the template stars), while
the velocity in \siii\ $\lambda$9069 is 2822$\pm$1 \kms.

Over the 6 arcseconds ($\sim$1 kpc) west of cluster \#23 
(position +4\arcsec), a shallow gradient of 40 \kms\ suggests a solid
body rotation curve centred at $\sim$+6\arcsec.
Similarly, while the two-dimensional \ha\ velocity field \citep[Figure
\ref{f-eso338-cigale} and ][]{o99} is far from ordered, a  
kiloparsec-scale region just west of the centre is nevertheless
consistent with solid body rotation and with the data we present
here. If interpreted as solid body rotation with the same inclination
(55\degr) indicated by the axis ratio of the outer isophotes, a
rotational mass of the order of $3\times 10^7$
\msun\ is implied. This is significantly smaller than the stellar mass
of the central region, which is at least $2\times 10^8$ \msun\
\citep{o01}.  The velocity dispersion of stars and gas in this region
is on the other hand 50 \kms\ or more, which indicates a dynamical
mass in excess of $10^9$ \msun\ if we can assume dynamical
equilibrium.

The slit passes through a number of the young star clusters identified
by \citet{OBR} (Figure \ref{f-eso338-slits-Ha}).  Of these, only at
the very massive cluster \#23 (at +4\arcsec), where the gas rotation
curves take a narrow excursion to the red, do we see any possible
effect on the stellar velocities.  This position coincides with the
maximum of the galaxy's continuum flux distribution and although
cluster \#23 accounts for up to half the continuum flux there, a
number of other clusters contribute to the signal at this position in
the slit. Here we also observe a gap in the \siii\ and
Paschen-line emission, consistent with the bubble-like distribution of
emission \citep{o07} seen in the HST \ha\ image. 

To investigate the source of the peak in the velocity curve, we
co-added the six two-dimensional spectra with the best seeing and
examined the structure around the \siii\ $\lambda9069$ line.  Inside
the gap at the position of \#23 there is a distinct, fainter emission
component with velocity about 2900 \kms, about 50 \kms\ redder than
the surrounding gas.  Emission components at $-$45 and +20 \kms\
relative to the stellar velocity of \#23 were observed by \citet{o07}
in \ha\ and \oiii\ in their slit passing through \#23 at PA 43$^\circ$. Our extracted
\siii\ profiles show signs of these components (Figure
\ref{f-eso338-ccf}). At +4\arcsec, the profile is distinctly
flat-topped. Inspection shows that the neighbouring profiles are
asymmetric with shallower red wings, giving the impression that a
redshifted component reaches its maximum at +4\arcsec, and is weaker
on either side. The velocity we measure in \caii\ at the position of
\#23 is identical within the errors to the cluster's velocity as
measured by \citet{o07} (2860$\pm$4
\kms), so it seems likely that we are seeing the same expanding bubble
around \#23 which \citet{o07} identified.

The velocity dispersion in the emission lines varies little along the
slit. It peaks at around 50 \kms\ at +4\arcsec\ and again at
+8\arcsec. These positions correspond to minima between the three
main concentrations of line emission along the slit. The stellar
velocity dispersions are larger than those of the gas: about 70$\pm$20
\kms\ compared to 40$\pm$10 \kms\ for the ionised gas at positions
where both are measured. We caution, however, that since the measured velocity
dispersions are comparable to the instrumental resolution (about 60
\kms), the errors in the velocity dispersion measurements may well be
underestimated.

In Figure \ref{f-eso338-cigale} we reproduce the spatially resolved
kinematics in the \ha\ line by \citet{o99}. We find good agreement
with our data presented here. The decline in velocity towards both
sides of the center at PA 90\degr-270\degr\ is well-reproduced, as is
the velocity difference along PA 160\degr-340\degr. Due to the lower
spatial resolution of the \ha\ data, the feature at cluster \#23
discussed above cannot be seen in Fig.~\ref{f-eso338-cigale}. On the
other hand, our slit spectra do not reach deep enough in the Paschen
lines to cover the tail toward the east where the velocity rises
again, or the peak in the velocity dispersion ($\sigma
\approx 60 \kms$) at 4\arcsec.5 east of where our two slits cross.
We do however see a local maximum in velocity dispersion in \siii\ at
position $-$5\arcsec\ (Figure \ref{f-rotcurve-eso338-1}).
For the slit at PA 160$^\circ$-340$^\circ$ (Figure
\ref{f-rotcurve-eso338-2}), the quality of the data is much poorer.
In the emission lines, we confirm the velocity gradient between
+1\arcsec\ and +4\arcsec\ noted by \citet{o99}. \citet{o01} 
argued that this gradient is unlikely to be due to rotation, and is
better explained by an outflow or a distinct dynamical component.
The velocity curve is symmetric about position +2\arcsec, and flat
beyond 0\arcsec\ on one side and +4\arcsec\ on the other, with a
spread in velocity of 40 \kms.  The velocity dispersion in the gas is
constant at $\sim$30 \kms\ across the slit.  The calcium triplet lines
are detected at at most two positions, corresponding to the eastern
end of the galaxy's central concentration at +1\arcsec, and the
vicinity of cluster \#25 \citep[from the outer sample of][]{OBR}. The
gas and stars show no differences at either of these positions, though
the stellar velocity dispersion at cluster \#25 is higher than that
for the gas.

We have also applied our current improved analysis methods to the
\esog\ spectra presented in \citepalias{o04}. We find no significant 
changes in either the results or conclusions of \citetalias{o04}.
\section{Discussion}

\subsection{The dynamical history of \esoig}

Earlier studies have suggested mergers and interactions as the
triggering mechanism of starbursts in luminous BCGs like \esoig\
\citep{o01,OB02}. Neutral hydrogen mapping of \esoig\ and its
companion \esoig\,B \citep{Cannon} showed evidence of a $\sim2\times10^9$-\msun\ tidal tail
stretching from just above the companion \esoig\,B to \esoig\ and beyond, indicating a strong
interaction in the recent past.  However, interaction with the
companion is unlikely to have caused the {\em present} burst of star
formation which was initiated about 40 Myr ago \citep{OZBR}. With a
projected distance between the two galaxies of 70 kpc, such timing
requires an extreme collision velocity of at least 1500
\kms. Moreover, the radial velocity difference of the two galaxies is
only a few $\times$10 \kms, and the companion is dynamically
well-behaved, both in \ha\ \citep{o01} and in \hi\ 21~cm
\citep{Cannon}. It is therefore more likely that the current burst is
unrelated to \esoig\,B. For typical velocities of galaxy pairs and
triples, of the order of 100 \kms, the last close passage of the two galaxies would have occurred
on the order of 1 Gyr ago. Interestingly, this corresponds to a peak
in the age distribution of the star clusters in \esoig\ \citep{OZBR}.

Based its morphology and its disturbed \ha\ velocity field,
\citet{o01} suggested a merger as the cause of the starburst in
\esoig. However, in addition to gravity, the interstellar medium in
galaxies can also be affected by feedback from stellar winds and
supernova explosions.  For this reason, a chaotic \ha\ velocity field
alone is not necessarily proof of a merger. An independent measurement
of the stellar and gas kinematics is required to separate the effects
of gravity and feedback.

In \esoig, we see that the velocity fields of the ionised gas and the
young stellar population tend to follow each other. This
demonstrates that this galaxy's chaotic \ha\ velocity field cannot be
explained solely by feedback in the form of expanding bubbles and
superwinds. Such processes certainly affect the velocity field, as
witnessed by the bubbles along slit PA 160$^\circ$-340$^\circ$ and
around cluster
\#23 \citep[see above and][]{o07}. But the velocity
curve traced in \caii\ is dynamically just as surprising as that of
the gas.

Can we be sure that the \caii\ triplet is a reliable tracer of the
stellar motions in the galaxy?  The triplet lines become observable as
soon as a stellar population is old enough to produce red giants and
supergiants, \ie\ after about 5 Myr, and by virtue of the strong
starburst in \esoig\ the \caii\ triplet absorption is likely dominated
by young supergiants.  This is evident in our data, as we see strong
\caii\ lines even where young super star clusters (SSCs) dominate.
There is of course the possibility that the young population dominates
the observed \caii\ velocity field, while the old stellar population
is still kinematically well-behaved. This would require that the
young stars were formed when an external gas cloud fell into the
gravitational potential of the galaxy but did not have enough mass to
significantly distort it.
While we cannot be sure that this did not happen, it seems to us
more likely that
the \caii\ lines  do indeed trace the major part of the
stellar motions.  In such a case, neither the stars nor the
ionised gas are in dynamical equilibrium.

An additional clue to \esoig's history comes from the tail-like structure
to the east of the starburst, observed in both \ha\ \citep{o99}
and in optical and near-IR broadband filters \citep{OB02}. It has
significantly bluer colours than the rest of the galaxy outside the
starburst region. Moreover, it contains young star clusters
\citep{OZBR} and clearly cannot solely be due to gas emission.
It is also coincident with a local maximum in the \hi\ column density
map \citep{Cannon}. \citet{o01} attributed this tail to a young,
off-centre stellar population with, again, peculiar kinematics.

Taken together, all these observations leave only one plausible
explanation for the origin of \esoig's starburst, morphology, globally
peculiar \ha, \hi\ and now its stellar kinematics.  It must be the
result of a recent merger, either with a dwarf galaxy or a massive
neutral gas cloud.  While we still do not know whether the stars are
as disrupted as the gas on larger scales, a merger is still required
to explain the disturbed central~kinematics.

Whether \esoig's merger was with another galaxy or with a gas cloud is
not a question we can answer given the current observations.  What 
triggers a starburst is the supply of gas to the central regions
\citep[\cf ][]{Combes}, and this could be provided either by an
infalling dwarf galaxy or massive cloud of gas. Indeed, large amounts
of neutral gas are still present close to the galaxy: in addition to
the tidal tail, the \hi\ data of \citet{Cannon} also indicate an
asymmetric distribution of neutral gas close to \esoig, most
concentrated on its west side. Moreover, the galaxy shows no 
evidence of a single nucleus, still less a double one, as might be
expected after a merger.

In either case, if its lack of rotational support on small scales is
also applicable at large galactocentric radii, \esoig\ may well be in
the process of evolving into an early-type galaxy.  Alternatively, the
lack of central rotational support for the gas and stars in the
starburst could be interpreted as the formation of a spherical or
triaxial system, or a bulge \citep{Hammer}.

\esoig\ shows a discrepancy between its
estimated stellar (photometric) mass and the dynamical mass inferred
from the \ha\ rotation curve \citep{o01}. Whereas a high dynamical
mass is usually interpreted as evidence for dark matter, here the 
galaxy is not rotating fast enough to support its stellar mass. \esog\ \citepalias{o04} is similar in this respect. In the case of
\esoig, \citet{o01} failed to derive a rotation curve where
the receding and approaching sides agreed.  
The two-dimensional velocity field is clearly very
irregular and suggests that the system is not in dynamical equilibrium
(\cf\ Figure~\ref{f-eso338-cigale}).

\subsection{Stellar and ionised gas kinematics in blue compact galaxies}

Adding our results to those of \citetalias{o04} and \citet{m07} for
He\,2-10, we have now studied the stellar kinematics of three blue
compact galaxies (BCGs) with disturbed velocity fields.  Here we
assess what the interim results of our programme have to say about
BCGs as a class, both locally and at higher redshift.

We have now seen two cases where the gas and stars are clearly
decoupled (He\,2-10 and \esog) and one where they for the most part
follow one another (\esoig).  In \esoig, moreover, we see evidence
that the coupling between gas and stars can vary depending on position
in the galaxy.

Our data has allowed us to identify three physical mechanisms
that contribute to the dynamics of these galaxies. 

\noindent {\em Gas-star decoupling:} The stellar and gaseous
velocity fields differ from each other, due either to outflows
triggered by supernova winds, or to infall in a merging process.  We
see this in both \esog\ \citepalias{o04} and He\,2-10 \citep{m07}.

\noindent {\em Large-scale gravitational disturbances:} The 
stellar velocity field is comparable to that of the ionised gas, and
neither the stars nor the gas may trace the potential.  This situation
can indicate real dynamical disturbances of the type expected in
mergers, and is what we observe in the central region of
\esoig.
 
\noindent {\em Velocity dispersion provides support:}  Stellar velocity 
dispersions in late-type galaxies are usually small, but in galaxies
which apparently lack rotational support, the velocity dispersion must
contribute to the gravitational support \citep*[\eg ][]{Cote}.  This
could explain at least the low rotational velocities in He\,2-10
and \esoig, if not the shapes of their rotation curves, and is
consistent with the observed emission-line widths and stellar velocity
dispersions \citep{o01}. Elliptical-like galaxies could, it seems, be
presently forming in these galaxies as a result of earlier mergers.

What are the implications for studies of similar galaxies at
intermediate redshift \citep[\eg][]{Bershady,Puech}
Observations of local late-type galaxies \citep{kg00} and
ultraluminous infrared galaxies \citep{Colina} have shown that stellar
and emission-line velocity amplitudes and velocity dispersions are
comparable.  Our BCG measurements show that the stellar velocity
dispersion is higher than that of the gas in \esoig, and in He\,2-10
the stellar velocity dispersion was lower than that of the gas in the
centre of the galaxy \citep{m07}. So far, it seems that emission-line
velocity fields in BCGs are not reliable tracers of the underlying
stellar velocities. An extended study of the combined stellar and gas
kinematics of a larger number local BCGs and other analogues of the
compact narrow emission line galaxies at intermediate redshift is
therefore required.

\section{Conclusions}

We have investigated the stellar velocity fields of the blue compact
galaxy \esoig, using the infrared \caii\ triplet.  Using
both cross-correlation with template stars and penalised pixel-fitting
to determine velocities and velocity dispersions, we show that stars
and gas can follow one another in these galaxies, but do not always do
so.

Our new results exclude the possibility that the chaotic kinematics
seen in the \ha\ gas in the centre is simply the result of feedback
and outflows, although such are probably present as well.  We conclude
that the centre of the galaxy is not rotationally supported, and that
the current starburst was most likely triggered by a merger,
either with a dwarf galaxy or a massive gas cloud.

\begin{acknowledgements}
We acknowledge helpful discussions with Javier Cenarro, Matthew Hayes,
Genoveva Micheva and Emanuela Pompei.  K.\ Fathi acknowledges support
from the Instituto de Astrof{\'\i}sica de Canarias through project P3/86,
the Royal Swedish Academy of Sciences' Hierta-Retzius foundation, and
the Wenner-Gren Foundation. N.\ Bergvall and G.\ {\"O}stlin acknowledge
support from the Swedish Research Council and the Swedish National Space
Board.  Figures in this
paper were prepared using the Perl Data Language. 
\end{acknowledgements}

\bibliographystyle{aa}
\bibliography{paper}

\begin{thebibliography}{42}
\expandafter\ifx\csname natexlab\endcsname\relax\def\natexlab#1{#1}\fi

\bibitem[{{Bergvall} \& {{\"O}stlin}(2002)}]{OB02}
{Bergvall}, N. \& {{\"O}stlin}, G. 2002, \aap, 390, 891

\bibitem[{{Bershady} {et~al.}(2005){Bershady}, {Vils}, {Hoyos}, {Guzm{\'a}n},
  \& {Koo}}]{Bershady}
{Bershady}, M.~A., {Vils}, M., {Hoyos}, C., {Guzm{\'a}n}, R., \& {Koo}, D.~C.
  2005, in Astrophysics and Space Science Library, Vol. 329, Starbursts: From
  30 Doradus to Lyman Break Galaxies, ed. R.~{de Grijs} \& R.~M. {Gonz{\'a}lez
  Delgado}, 177--+

\bibitem[{{Cannon} {et~al.}(2004){Cannon}, {Skillman}, {Kunth}, {Leitherer},
  {Mas-Hesse}, {Östlin}, \& {Petrosian}}]{Cannon}
{Cannon}, J.~M., {Skillman}, E.~D., {Kunth}, D., {et~al.} 2004, \apj, 608, 768

\bibitem[{{Cappellari} \& {Emsellem}(2004)}]{CE04}
{Cappellari}, M. \& {Emsellem}, E. 2004, \pasp, 116, 138

\bibitem[{{Cenarro} {et~al.}(2001{\natexlab{a}}){Cenarro}, {Cardiel}, {Gorgas},
  {Peletier}, {Vazdekis}, \& {Prada}}]{Cenarro01a}
{Cenarro}, A.~J., {Cardiel}, N., {Gorgas}, J., {et~al.} 2001{\natexlab{a}},
  \mnras, 326, 959

\bibitem[{{Cenarro} {et~al.}(2001{\natexlab{b}}){Cenarro}, {Gorgas}, {Cardiel},
  {Pedraz}, {Peletier}, \& {Vazdekis}}]{Cenarro01b}
{Cenarro}, A.~J., {Gorgas}, J., {Cardiel}, N., {et~al.} 2001{\natexlab{b}},
  \mnras, 326, 981

\bibitem[{{Colina} {et~al.}(2005){Colina}, {Arribas}, \&
  {Monreal-Ibero}}]{Colina}
{Colina}, L., {Arribas}, S., \& {Monreal-Ibero}, A. 2005, \apj, 621, 725

\bibitem[{{Combes}(2005)}]{Combes}
{Combes}, F. 2005, in AIP Conference Proceedings, Vol. 783, The Evolution of
  Starbursts: The 331st Wilhelm and Else Heraeus Seminar, 43--49

\bibitem[{{C{\^o}t{\'e}} {et~al.}(2000){C{\^o}t{\'e}}, {Carignan}, \&
  {Freeman}}]{Cote}
{C{\^o}t{\'e}}, S., {Carignan}, C., \& {Freeman}, K.~C. 2000, \aj, 120, 3027

\bibitem[{{Garland} {et~al.}(2007){Garland}, {Pisano}, {Williams}, {Guzman},
  {Castander}, \& {Sage}}]{Garland}
{Garland}, C.~A., {Pisano}, D.~J., {Williams}, J.~P., {et~al.} 2007, ArXiv
  e-prints, 708

\bibitem[{{Gil de Paz} {et~al.}(1999){Gil de Paz}, {Zamorano}, \&
  {Gallego}}]{gdp}
{Gil de Paz}, A., {Zamorano}, J., \& {Gallego}, J. 1999, \mnras, 306, 975

\bibitem[{{Guzman} {et~al.}(1997){Guzman}, {Gallego}, {Koo}, {Phillips},
  {Lowenthal}, {Faber}, {Illingworth}, \& {Vogt}}]{g97}
{Guzman}, R., {Gallego}, J., {Koo}, D.~C., {et~al.} 1997, \apj, 489, 559

\bibitem[{{Guzman} {et~al.}(1996){Guzman}, {Koo}, {Faber}, {Illingworth},
  {Takamiya}, {Kron}, \& {Bershady}}]{g96}
{Guzman}, R., {Koo}, D.~C., {Faber}, S.~M., {et~al.} 1996, \apj, 460, 5

\bibitem[{{Hammer} {et~al.}(2005){Hammer}, {Flores}, {Elbaz}, {Zheng}, {Liang},
  \& {Cesarsky}}]{Hammer}
{Hammer}, F., {Flores}, H., {Elbaz}, D., {et~al.} 2005, \aap, 430, 115

\bibitem[{{Hayes}(2007)}]{Hayes}
{Hayes}, M. 2007, PhD thesis, Stockholm University

\bibitem[{{Ho} \& {Filippenko}(1996{\natexlab{a}})}]{HFa}
{Ho}, L.~C. \& {Filippenko}, A.~V. 1996{\natexlab{a}}, \apjl, 466, L83+

\bibitem[{{Ho} \& {Filippenko}(1996{\natexlab{b}})}]{HFb}
{Ho}, L.~C. \& {Filippenko}, A.~V. 1996{\natexlab{b}}, \apj, 472, 600

\bibitem[{{Jesseit} {et~al.}(2007){Jesseit}, {Naab}, {Peletier}, \&
  {Burkert}}]{Jesseit}
{Jesseit}, R., {Naab}, T., {Peletier}, R.~F., \& {Burkert}, A. 2007, \mnras,
  376, 997

\bibitem[{{Kobulnicky} \& {Gebhardt}(2000)}]{kg00}
{Kobulnicky}, H.~A. \& {Gebhardt}, K. 2000, \aj, 119, 1608

\bibitem[{{Koo} {et~al.}(1995){Koo}, {Guzman}, {Faber}, {Illingworth},
  {Bershady}, {Kron}, \& {Takamiya}}]{koo}
{Koo}, D.~C., {Guzman}, R., {Faber}, S.~M., {et~al.} 1995, \apj, 440, L49

\bibitem[{{Kronberger} {et~al.}(2007){Kronberger}, {Kapferer}, {Schindler}, \&
  {Ziegler}}]{Kronberger}
{Kronberger}, T., {Kapferer}, W., {Schindler}, S., \& {Ziegler}, B.~L. 2007,
  \aap, 473, 761

\bibitem[{{Marquart} {et~al.}(2007){Marquart}, {Fathi}, {{\"O}stlin},
  {Bergvall}, {Cumming}, \& {Amram}}]{m07}
{Marquart}, T., {Fathi}, K., {{\"O}stlin}, G., {et~al.} 2007, \aap, 474, L9

\bibitem[{{Matheson} {et~al.}(2000){Matheson}, {Filippenko}, {Ho}, {Barth}, \&
  {Leonard}}]{m00}
{Matheson}, T., {Filippenko}, A.~V., {Ho}, L.~C., {Barth}, A.~J., \& {Leonard},
  D.~C. 2000, \aj, 120, 1499

\bibitem[{{Melnick} {et~al.}(1987){Melnick}, {Moles}, {Terlevich}, \&
  {Garcia-Pelayo}}]{melnick87}
{Melnick}, J., {Moles}, M., {Terlevich}, R., \& {Garcia-Pelayo}, J.-M. 1987,
  \mnras, 226, 849

\bibitem[{{Nelson} \& {Whittle}(1995)}]{nw95}
{Nelson}, C.~H. \& {Whittle}, M. 1995, \apjs, 99, 67

\bibitem[{{Ochsenbein} {et~al.}(2000){Ochsenbein}, {Bauer}, \&
  {Marcout}}]{VizieR}
{Ochsenbein}, F., {Bauer}, P., \& {Marcout}, J. 2000, \aaps, 143, 23

\bibitem[{Osterbrock(1989)}]{o89}
Osterbrock, D.~E. 1989, {Astrophysics of gaseous nebulae and active galactic
  nuclei} (University Science Books)

\bibitem[{{Osterbrock} {et~al.}(1996){Osterbrock}, {Fulbright}, {Martel},
  {Keane}, {Trager}, \& {Basri}}]{o96}
{Osterbrock}, D.~E., {Fulbright}, J.~P., {Martel}, A.~R., {et~al.} 1996, \pasp,
  108, 277

\bibitem[{{Osterbrock} \& {Martel}(1992)}]{o92}
{Osterbrock}, D.~E. \& {Martel}, A. 1992, \pasp, 104, 76

\bibitem[{{{\"O}stlin} {et~al.}(2001){{\"O}stlin}, {Amram}, {Bergvall},
  {Masegosa}, {Boulesteix}, \& {M{\'a}rquez}}]{o01}
{{\"O}stlin}, G., {Amram}, P., {Bergvall}, N., {et~al.} 2001, \aap, 374, 800

\bibitem[{{{\"O}stlin} {et~al.}(1999){{\"O}stlin}, {Amram}, {Masegosa},
  {Bergvall}, \& {Boulesteix}}]{o99}
{{\"O}stlin}, G., {Amram}, P., {Masegosa}, J., {Bergvall}, N., \& {Boulesteix},
  J. 1999, \aaps, 137, 419

\bibitem[{{{\"O}stlin} {et~al.}(1998){{\"O}stlin}, {Bergvall}, \&
  {R\"onnback}}]{OBR}
{{\"O}stlin}, G., {Bergvall}, N., \& {R\"onnback}, J. 1998, \aap, 335, 85

\bibitem[{{{\"O}stlin} {et~al.}(2004){{\"O}stlin}, {Cumming}, {Amram},
  {Bergvall}, {Kunth}, {M{\'a}rquez}, {Masegosa}, \& {Zackrisson}}]{o04}
{{\"O}stlin}, G., {Cumming}, R.~J., {Amram}, P., {et~al.} 2004, \aap, 419, L43,
  (Paper I)

\bibitem[{{{\"O}stlin} {et~al.}(2007){{\"O}stlin}, {Cumming}, \&
  {Bergvall}}]{o07}
{{\"O}stlin}, G., {Cumming}, R.~J., \& {Bergvall}, N. 2007, \aap, 461, 471

\bibitem[{{{\"O}stlin} {et~al.}(2003){{\"O}stlin}, {Zackrisson}, {Bergvall}, \&
  {R{\"o}nnback}}]{OZBR}
{{\"O}stlin}, G., {Zackrisson}, E., {Bergvall}, N., \& {R{\"o}nnback}, J. 2003,
  \aap, 408, 887

\bibitem[{{Pisano} {et~al.}(2001){Pisano}, {Kobulnicky}, {Guzm{\'a}n},
  {Gallego}, \& {Bershady}}]{Pisano}
{Pisano}, D.~J., {Kobulnicky}, H.~A., {Guzm{\'a}n}, R., {Gallego}, J., \&
  {Bershady}, M.~A. 2001, \aj, 122, 1194

\bibitem[{{Press} {et~al.}(1992){Press}, {Teukolsky}, {Vetterling}, \&
  {Flannery}}]{press}
{Press}, W.~H., {Teukolsky}, S.~A., {Vetterling}, W.~T., \& {Flannery}, B.~P.
  1992, {Numerical recipes in {\sc FORTRAN}. The art of scientific computing}
  (Cambridge: University Press)

\bibitem[{{Puech} {et~al.}(2006){Puech}, {Hammer}, {Flores}, {{\"O}stlin}, \&
  {Marquart}}]{Puech}
{Puech}, M., {Hammer}, F., {Flores}, H., {{\"O}stlin}, G., \& {Marquart}, T.
  2006, \aap, 455, 119

\bibitem[{{Terlevich} \& {Melnick}(1981)}]{tm81}
{Terlevich}, R. \& {Melnick}, J. 1981, \mnras, 195, 839

\bibitem[{{Tonry} \& {Davis}(1979)}]{TD}
{Tonry}, J. \& {Davis}, M. 1979, \aj, 84, 1511

\bibitem[{{Wegner} {et~al.}(1999){Wegner}, {Colless}, {Saglia}, {McMahan},
  {Davies}, {Burstein}, \& {Baggley}}]{Wegner99}
{Wegner}, G., {Colless}, M., {Saglia}, R.~P., {et~al.} 1999, \mnras, 305, 259

\bibitem[{{Werk} {et~al.}(2004){Werk}, {Jangren}, \& {Salzer}}]{Werk}
{Werk}, J.~K., {Jangren}, A., \& {Salzer}, J.~J. 2004, \apj, 617, 1004

\end{thebibliography}

\end{document}